\documentclass[conference]{IEEEtran}
\IEEEoverridecommandlockouts
\usepackage{cite}
\usepackage{amsmath,amssymb,amsfonts}
\usepackage{algorithmic}
\usepackage{graphicx}
\usepackage{subfigure}
\usepackage{textcomp}
\usepackage{xcolor}
\usepackage{color}
\usepackage{multirow}

\newcommand{\customcomment}[3]{\textcolor{#1}{#3}}
\newcommand{\wen}[1]{\customcomment{black}{w}{#1}}

\def\BibTeX{{\rm B\kern-.05em{\sc i\kern-.025em b}\kern-.08em
    T\kern-.1667em\lower.7ex\hbox{E}\kern-.125emX}}
\begin{document}

\title{
Improving Conversational Recommender System by Pretraining Billion-scale Knowledge Graph}
\author{
\IEEEauthorblockN{Chi-Man Wong\IEEEauthorrefmark{3}\IEEEauthorrefmark{2},Fan Feng\IEEEauthorrefmark{3}, 
Wen Zhang\IEEEauthorrefmark{4}, 
Chi-Man Vong\IEEEauthorrefmark{2},
Hui Chen\IEEEauthorrefmark{3}, Yichi Zhang\IEEEauthorrefmark{3}, Peng He\IEEEauthorrefmark{3},\\ Huan Chen\IEEEauthorrefmark{1}, Kun Zhao\IEEEauthorrefmark{3}, 
Huajun Chen\IEEEauthorrefmark{4}\IEEEauthorrefmark{5}
}
\IEEEauthorblockA{\IEEEauthorrefmark{3}\IEEEauthorrefmark{1}Alibaba Group, \IEEEauthorrefmark{2}University of Macau, \IEEEauthorrefmark{4}Zhejiang University, China} 

}\maketitle


\begin{abstract}
Conversational Recommender Systems (CRSs) in E-commerce \wen{platforms} aim to recommend 
items to users via multiple conversational interactions. 
\wen{Click-through rate (CTR) prediction models are commonly used for ranking candidate items.}
However, 
\wen{most} 
CRSs are \wen{suffer from the problem of data scarcity and sparseness.}
\wen{To address this issue,}
 we propose a novel knowledge-enhanced deep cross network (K-DCN), a two-step (pretrain and fine-tune) CTR prediction model to recommend items\wen{.} 
We first construct a 
\wen{billion-scale}
conversation knowledge graph \wen{(CKG)} from 
\wen{ information about users, items and converations, }
and \wen{then} pretrain 
\wen{CKG} by introducing 
\wen{knowledge graph embedding method and graph convolution network to encode semantic and structural information respectively.}
\wen{To make the CTR prediction model sensible of current state of users and the relationship between dialogues and items, we introduce user-state and dialogue-interaction representations based on pre-trained CKG and propose K-DCN.}
\wen{In K-DCN, we fuse the user-state representation, dialogue-interaction representation and other normal feature representations via deep cross network, which will give the rank of candidate items to be recommended.}
\wen{We experimentally prove that our proposal significantly outperforms baselines and show it's real application in Alime.}
\end{abstract}


\section{Introduction}
\begin{figure}[t]
        \center{\includegraphics[width=8cm,height=4cm]
        {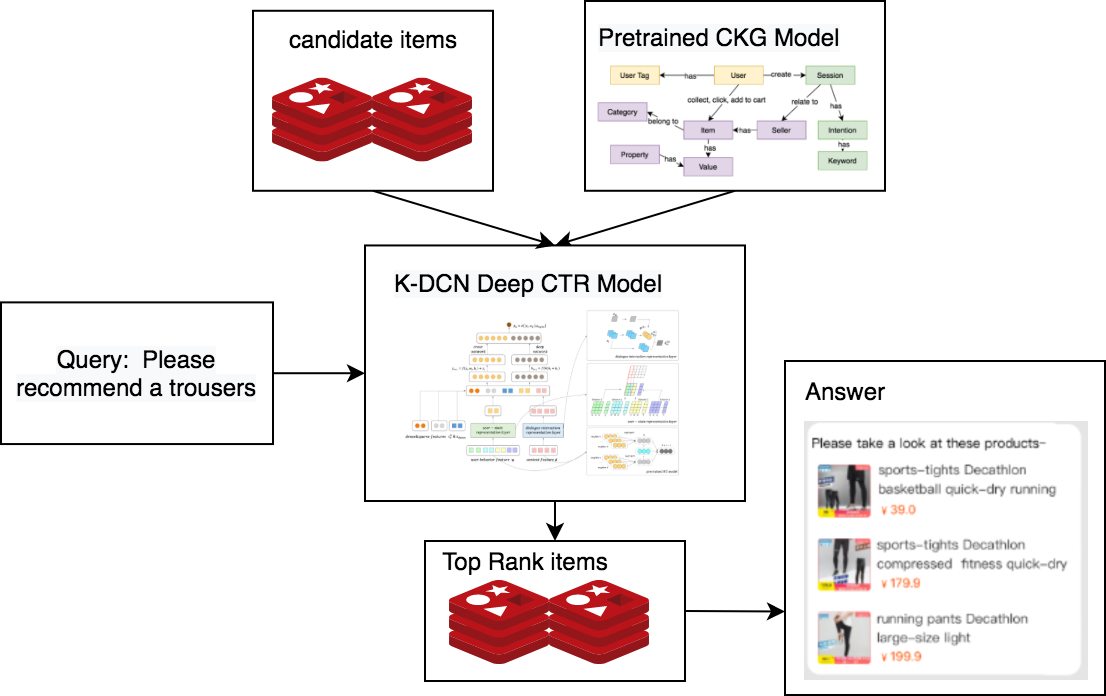}}
        \caption{\label{fig:task}  The workflow of K-DCN.}
        \label{fig:workflow} 
\end{figure}

Conversational Recommender Systems \wen{(CRS)}\cite{gao2018neural,rahman2017programming,fadhil2018can,nica2018chatbot,hildebrand2019ai} are commonly used to improve 
\wen{customer experience on E-commerce platforms.}
Many online stores have 
customer service chatbots
\wen{to help users find their ideal items, which   contributes to the revenue greatly.}
The goal of these chatbot\wen{s} is to identify 
\wen{users'}
intentions by multiple 
\wen{conversational}
interaction\wen{s}, and then 
\wen{recommend}
a list of most suitable items to them. 
As shown in figure \ref{fig:workflow},
in a certain dialogue, when \wen{a} user ask\wen{s} for recommendation, the CRS employs a three-stage pipeline. 
\wen{First, it makes semantic analysis of the query, then gets a list of candidate items, and finally  ranks candidate items through deep click-through rate (CTR) models.}

In this paper, we focus on the ranking \wen{process of CRS}, thus the key task is to estimate the
\wen{CTR}.
\wen{CTR prediction models\cite{wang2017deep,guo2017deepfm,wang2018dkn} with various deep networks have been proposed and achieved good performance.}
However, 
\wen{they}
mainly rely on abundant behavioral records of users on items, 
\wen{while}
these records might be very sparse because 
\wen{of} 
limited interaction\wen{s} with a few chatbots among millions of online 
\wen{shops}, which 
would lead to the over-fitting problem. 

\wen{To solve the data scarcity and sparseness problem, we propose to make user, item and conversation information into consideration, since certain preference for items might be expressed in conversation. 
}

\wen{To better organize and utilize information, our proposal mainly includes two parts: 1) construct and pre-train of conversational knowledge graph (CKG); and 2) knowledge-enhanced deep cross network (K-DCN) is employed for fine-tuning. 
Firstly, we construct a billion-scale CKG from information about user, item and  conversation, and then is employed to learn a pre-trained model to obtain better entity representations. There are two common approaches to learn entity representations in KGs. The one is Graph Neural Network(GNN) based methods\cite{kipf2016semi,li2018adaptive,zhuang2018dual,velivckovic2017graph,zhang2018gaan} like GCN~ \cite{kipf2016semi}, which models structure information of an entity by aggregating all its neighbors' information. 
The other one is knowledge graph emebdding (KGE) methods like translation-based methods\cite{bordes2013translating,wang2014knowledge,lin2015learning,ji2015knowledge,sun2019transedge,yang2014embedding, sun2019rotate,liu2019geniepath,CrossE}, which are good at capture semantic information of entities. Since both structure and semantic information of entities are important, we combine GNN-based method and KGE method together for pre-training CKG via joint learning.  
In the second part,
we propose a novel two-step ranking model, K-DCN, in which we introduce 
user-state representations and dialogue-interaction representations based on pre-trained CKG, to help the CTR prediction model sensible of the uses' state and relationships between dialogues and items. We experiment our proposal on real-life e-commerce CRS. The results show that our proposal outperforms baselines and converge faster during training. Finally, we also show the application of our proposal in Alimi, a chatbot in Taobao platform. }

\wen{In summary,} our contributions are: 
\begin{itemize}
    \item  \wen{We propose to integrate both user, item and conversation information in the form of conversation knowledge graph to solve the data scarcity and sparseness problem of CTR prediction models in CRS.}
    \item 
    \wen{We introduce a novel method to organize and utilize various information, via first get good entity representations from billion-scale pre-trained CKG, and then fine-tune K-DCN with user-state representations and dialogue-interaction representation introduced. }
    \item 
    \wen{We experimentally show that our proposal outperforms baselines and converge faster during training on real-life e-commerce datasets, and we also show the real application of our proposal on Alimi.} 
\end{itemize}

\section{Related Work}

\subsection{Translation-based KG embedding}
Translation-based methods adopt a scoring function $f(h,r,t)$ to measure the plausibility of a fact \wen{(h,r,t)} from KG. E.g., in 
TransE \cite{bordes2013translating} 
\wen{assumes $f(h,r,t) = \|\mathbf{h} + \mathbf{r} - \mathbf{t}\|$}.
TransH \cite{wang2014knowledge} 
\wen{projects entity representation onto relation' hyperplane before calculating score.}
TransR \cite{lin2015learning} and TransD \cite{ji2015knowledge}
\wen{project entities from entity space to relation space via projection matrix.}
TransEdge \cite{sun2019transedge} contextualizes relation by edge-centric embedding.
DistMult \cite{yang2014embedding} captures relational semantics by the composition of relation; RotateE \cite{sun2019rotate} learns various relation patterns by modeling relation as rotation. However, all of the above methods do not learn the structure information of KG, which is an important information to provide robustness of the model.

\subsection{GCN-based KG embedding}
GCN-based method represents the embedding of each entity by iteratively propagating neighbor information. For example, GCN \cite{kipf2016semi} introduces a first-order approximation of ChebNet to perform graph convolution so that the number of parameters are restrained and the issue of over-fitting is avoided; 
AGCN \cite{li2018adaptive} could learn all graph structure information by introducing a distance metric learning; 
DGCN \cite{zhuang2018dual}propose a dual graph convolution to encodes both local and global structure information by normalized adjacency and positive pointwise mutual information (PPMI) matrix; GAT \cite{velivckovic2017graph} employed masked self-attentional layer to learn the importance of neighborhoods' information by assigning different weights to them;  GeniePath \cite{liu2019geniepath} learns the importance of different sized neighborhoods by using an adaptive path layer, such that both breadth and depth can be explored for information extraction; Gaan  \cite{zhang2018gaan} control each attention head's importance by sub-convolution network; Although these methods could encode the neighbor's information of an entity, the semantic meaning can not be learned, which would deteriorate the performance.

\subsection{Deep Click-through Rate Prediction model}
The development of deep neural networks (DNNs) and embedding techniques 
provides a better way to learn feature expression in recommender systems.  For example,
Deep \& Cross Network(DCN) \cite{wang2017deep} and DeepFM \cite{guo2017deepfm,wang2020research} 
contain both deep component and shallow component to capture the feature 
interaction automatically. In common Click-through Rate Prediction
model, the input data with sparse and dense features mainly 
concern about user action and product attribute, and only use the embedding technique
to reduce the dimensionality of categorical features. But in dialog system, 
the input must be more specific about current status. Inspired by the DKN
\cite{wang2018dkn}, the knowledge-level embeddings of entities on the 
dialog system can enrich the use state and dialogue interaction representation directly.

  \begin{figure*}[ht]
        \center{\includegraphics[width=13cm,height=7cm]
        {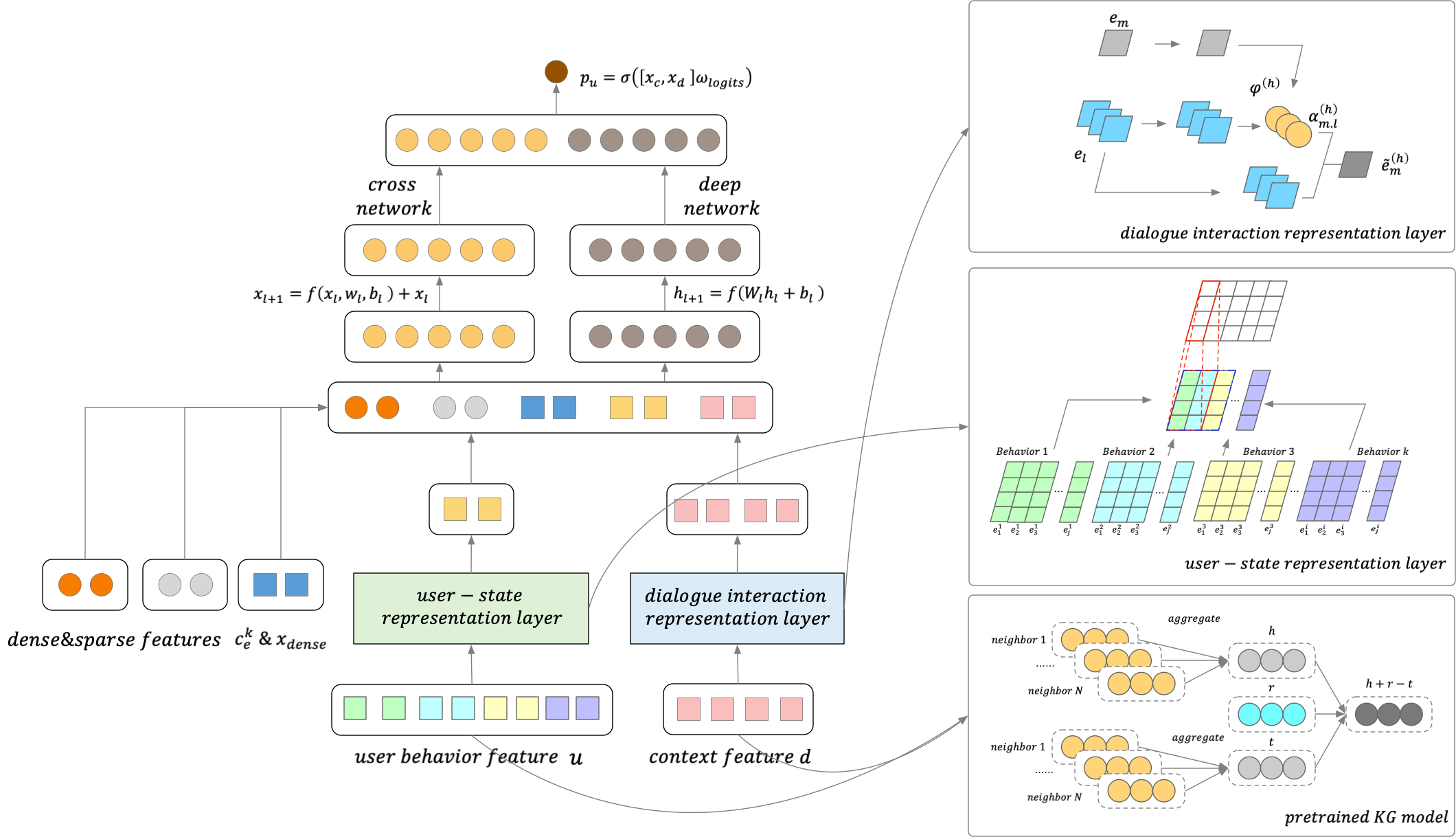}}
        \caption{\label{fig:model} The architecture of knowledge-enhanced deep cross network (K-DCN). }
        \label{figmodel}
      \end{figure*}
      

\begin{figure}[!hbpt]
        \center{\includegraphics[width=8cm,height=4cm]
        {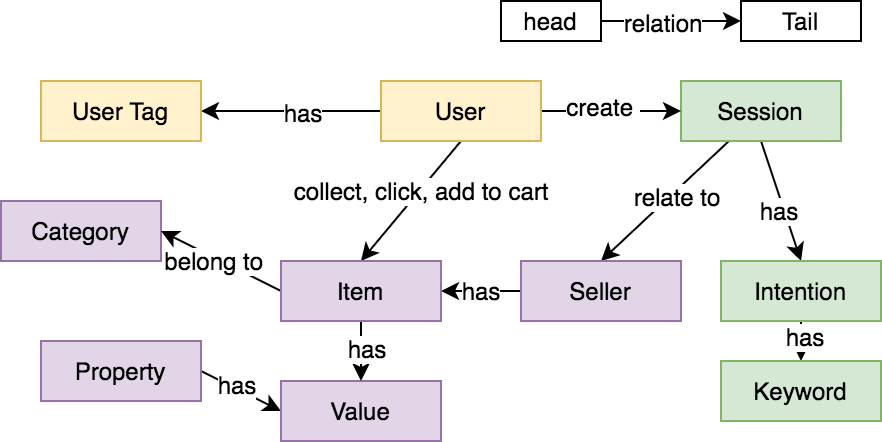}}
        \caption{\label{fig:task}  The conversation knowledge graph.}
        \label{fig:kg} 
\end{figure}

\section{Method}

Our method contains two parts: 1) constructing a Conversation Knowledge Graph(CKG) and pre-training it to encode structure and semantic information; 2) fine-tuning a ranking model DCN based on normal features and features from pre-trained knowledge graph representations, including user-state representations and dialogue-interaction representations.

\subsection{Constructing and Pre-training of CKG}
\subsubsection{Conversation  Knowledge Graph Construction}
We make knowledge graph as a way to encode diverse information related to conversational recommendation system and we construct the CKG from real-world scenarios of user-chatbot conversation. 
There are three kinds of information contained in CKG:
\begin{itemize}
    \item Information   about users: in our platform, each user has a list of tags, such as gender, shopping history, etc, which could help to catch the interest of users. Thus we build triples encoding the tag information in CKG in the form of (user, has, user tag). 
    \item Information  about items: our platform contains rich information about items like category and seller of items, and also properties of items such as color, brand, etc. For categories and seller information, we build triple in the form of (item, belong to, category) and (seller, has, item) respectively. For properties and  values, two kinds of triples are built (item, has, value) and (property, has, value). This  item information could contribute the mapping to users interests greatly. 
    \item Information  about conversation: in the chatbot, many sessions are created when user chat with the bot, the intentions and keywords from the conversation could help understand the intention of a user more precisely. Thus,we could build four kinds of triples based on conversation history, (user, created, session), (session, relate to, seller), (session, has, intention) and (intention, has, keywords). 
\end{itemize} 
The schema of CKG is shown in figure 3. Finally, CKG contains 600 million entities and 6 billion triples.


\subsubsection{Conversation Knowledge Graph Pre-training}
As known that structural and semantic information are both valuable in knowledge graph, thus our CKG pre-training
contains two modules,  structural embedding module and semantic embedding module to capture them respectively.
 
\emph{Structural Embedding Module:} 
Graph neural network(GNN) has been proved to be useful in encoding structural information\cite{kipf2016semi,li2018adaptive,zhuang2018dual,velivckovic2017graph,zhang2018gaan}. 
Thus we employ GCN \cite{kipf2016semi} to learn the structur representation of entities in CKG where a vector representation $\mathbf{e} \in \mathbb{R}^{d}$ for each entity $e \in \mathcal{K}$. 
Gathering all entity embedding together in order, we get $\mathbf{E} \in \mathbb{R}^{n_e \times d}$, where $n_e$ is the number of entities and $d$ is the embedding dimension.
A multi-layer GCN is given with a simple layer-wise propagation rule and the $i$th layer could be represented as:
\begin{equation}
X^{(i+1)} = \sigma (\hat{D}^{-\frac{1}{2}}\hat{A}\hat{D}^{-\frac{1}{2}}X^{(i)}W^{(i)}  )
\end{equation}
where $\hat{A} \in \mathbb{R}^{n\times n}$ is adjacency matrix where $\hat{A}_{ij}=1$ if $e_i$ and $e_j$ are connected in CKG and $\hat{A}_{ij}=0$ otherwise. $\hat{D}$ is the diagonal degree matrix of $\hat{A}$. $n$ is the number of entities in CKG.
$\sigma$ is the activation function and we make it sigmoid function during experiments. $W^{(i)}$ is the weight matrix in $i$th layer. The output $X^{(i+1)}$ encodes the structure information from $i$th layer with $X^{(i)}$ as structure information of previous layers. In the first layer, we make $X^{(1)} = \mathbf{E}$.

\emph{Semantic Embedding Module:}
Many knowledge graph embedding(KGE) methods are proposed to encode the semantic information implicitly. Considering the effectiveness and efficiency, we employ TransE in semantic embedding module.  For a triple $(e_i, r, e_j)$, the score function of it is defined as:
\begin{equation}\label{eq:1}
f(e_i, r, e_j) = ||\mathbf{e}_i +\mathbf{r} - \mathbf{e}_j||
\end{equation}
where $\mathbf{e}_i, \mathbf{r}$ and $\mathbf{e}_j$ are embeddings of $e_i, r$ and $e_j$ respectively. With this function, positive triples should get lower scores and negative ones get higher ones.

\emph{Pretraining:}
we jointly train structure embedding module and semantic embedding module to learn entity and relation embeddings.
During co-training, 
we make the output of multi-layer GCN $X^{(i+1)}$ as the entity embedding matrix for TransE which means $\mathbf{e}_k = X_k^{(i+1)} (k \in [0, n-1])$. The training target to minimize is the following margin-based ranking loss:
\begin{equation}
     L = \sum_{(e_i,r,e_j) \in \mathcal{T}} [f(e_i,r,e_j) + \gamma - f(e_i^\prime, r^\prime, e_j^\prime)]_+
\end{equation}
where $(e_i^\prime, r^\prime, e_j^\prime)$ is the negative sample of $(e_i,r,e_j)$ by randomly replace $e_i$
 or $e_j$ with $e\in CKG$. Function $[x]_{+}=0$ if $x<0$,  otherwise $[x]_{+}=x$. 
After construction and pre-training of CKG, the trained entity embeddings will be used to fine-tune CTR prediction models to improve conversational recommendation system.

\subsection{K-DCN for Fine-tuning}
Normally, common sparse and dense features, eg. statistical features, are used in CTR prediction models, while in conversation recommendation system, user's immediate question to chatbot is also important for  understanding user's intention, thus we consider and construct user-state and dialogue-interaction representations based on pre-trained CKG as additional features. 


\subsubsection{User-state Representation}
Users' state could be captured by their previous behaviors. For one user, suppose he/she has $k$ behaviors $\mathcal{B} = \{b_i | b_i = \{ e_1^{(i)}, e_2^{(i)} , ... \}, i\in [1,k] \}$ where $b_i$ is an user-item click sequence. For each $b_i$, we first average the embedding of each item $e \in b_i$ from pre-trained CKG and get a behavior vector $\mathbf{b}_i$:
\begin{equation}
    \mathbf{b}_i = \frac{1}{| b_i|} \sum_{e_j^{(i)} \in b_i} \mathbf{e}_j^{(i)}
\end{equation}


Then we vertically concatenate all behavior vectors as $\mathbf{B} \in \mathbb{R}^{d \times k}$ as shown in Figure (\ref{figmodel}) and employ convolutional neural networks(CNN)\cite{krizhevsky2012imagenet} on $\mathbf{B}$ to model the local information of the user-state representations:
\begin{equation}
    \mathbf{u}=f_{CNN}(\mathbf{B}*\mathbf{w} +b ) 
\end{equation}
where $*$ is is the convolution operator, and 
$b \in \mathbb{R}$ is a bias. We show the details of this step in right side of  Figure (\ref{figmodel}) marked as user-state representation layer.


\subsubsection{Dialogue-interaction Representation}
To build dialogue interactions, we first extract a set of keywords from users question $\mathcal{W}_{query}= \{w_q^{(1)}, w_q^{(2)}, ..., w_q^{(m)}\}$. Given a candidate item $c$, we also extract a set of keywords from its title $\mathcal{W}_{title}= \{w_t^{(1)}, w_t^{(2)}, ..., w_t^{(n)}\}$. Then, we gather all keywords together and get their representations from pre-trained CKG's entity embeddings, represented as $\mathbf{W} = \{\mathbf{w}^{(1)}, \mathbf{w}^{(2)}, ..., \mathbf{w}^{(m+n)}\}$. 
Then we 
fed keyword embeddings into a multi-head self attention network \cite{vaswani2017attention} to model the inter-relationship between query and candidate item. 
Formally:
\begin{equation}\label{eq:5}
   a_{ij} = \frac{exp((\mathbf{M}_a  \mathbf{w}_i) \circ (\mathbf{M}_b \mathbf{w}_j)))}{\sum _{t = 1} ^{m+n} exp((\mathbf{M}_a \mathbf{w}_i) \circ (\mathbf{M}_b \mathbf{w}_k))}
    \end{equation}
where $\circ$ is inner-product operation and $\mathbf{W}_a \in \mathbb{R}^{d \times d}$ and $\mathbf{W}_b \in \mathbb{R}^{d \times d}$ are matrices transforming input embeddings to a new space.
where $\psi(\mathbf{w}_{i},\mathbf{w}_{j}) = inner product({M_{a}\mathbf{w}_{i},{M}_{b}\mathbf{w}_{j}}) $ is the attention function.${M}_{a},{M}_{b}$ are matrix which  

Next, we update the representation through a weighted sum according to  $a_{ij}$:
\begin{equation}\label{eq:1}
\widetilde{\mathbf{w}}^{(i)} = \sum _ {j=1}^{m+n} a_{i,j} (\mathbf{W}_v \mathbf{e}^{(i)}) 
\end{equation}
Finally, we concatenate all updated word embeddings as the dialogue-interaction representation $\mathbf{d} \in \mathbb{R}^{(m+n)\times d }$:
\begin{equation}
    \mathbf{d} = [ \widetilde{\mathbf{w}}^{(1)},
    \widetilde{\mathbf{w}}^{(2)},
    ...,
    \widetilde{\mathbf{w}}^{(m+n)}]
\end{equation}


\subsubsection{K-DCN Model}
Inputs of K-DCN are various feature representations for a  candidate item $e$. Besides  aforementioned user-state and dialogue interaction representations $\mathbf{u}$ and $\mathbf{d}$, other common features are also considered,  including categorical features and statistical features. 
First, we concatenate all feature vectors as follows:
\begin{equation}
\mathbf{f} = [ \mathbf{c}_{e}^{(1)}, \mathbf{c}_{e}^{(2)}, ..., \mathbf{c}_{e}^{(k)}, \mathbf{x}_{dense}, \mathbf{u}, \mathbf{d}] 
\end{equation}
where $ \mathbf{c}_e^{(1)}, \mathbf{c}_e^{(2)}, ..., \mathbf{c}_e^k $ are embeddings of categories that target item $e$ belongs to, which will be randomly initialized during training.  
$\mathbf{x}_{dense}$ is statistical feature vector of $e$ where values could be price, sales etc.



  
With the feature representation $\mathbf{f}$ for $e$, we feed it into the cross network and deep network respectively which are two common components in CTR prediction model. 
The cross network is composed of $n_c$ cross layers and the $i$th layer could be represented as:
\begin{equation}
\mathbf{x}^{(i+1)}_c=\mathbf{f} (\mathbf{x}^{(i)}_c)^{T} \mathbf{w}_{i}+\mathbf{b}_{i}+\mathbf{x}^{(i)}_c
\end{equation}
where $\mathbf{x}_c^0 = \mathbf{f}$. 
The deep network is a $n_d$ layers of fully-connected feed-forward network and the function of the $i$th layer is
\begin{equation}
\mathbf{x}^{(i+1)}_d=\sigma\left(W_{i} \mathbf{x}^{(i)}_d +\mathbf{b}_{l}\right)
\end{equation}

Then we concatenate the two outputs from the above two networks
and feed the concatenated vector into a standard logits layer.
Formally:
\begin{equation}\label{eq:1}
  p=\sigma\left(\left[\mathbf{x}_{c}^{(n_c)}, \mathbf{x}_{d} ^{(n_d)}\right] \mathbf{W}_{\text {logits }}\right)
  \end{equation}
To train K-DCN, we minimize the following log likelihood loss function:
\begin{equation}
  L_{K-DCN}=-\frac{1}{N} \sum_{i=1}^{N} y_{i} \log \left(p_{i}\right)+\left(1-y_{i}\right) \log \left(1-p_{i}\right)
\end{equation}
where  $N$ is the number of input samples. $y_{i} \in \mathbb{R}^{n_{cand} \times 1}$ is the recommendation label of the $i$th sample and $n_{cand}$ is the number of candidate items to be recommended.


\begin{figure*}
\centering  
\subfigure[example A]{
\label{Fig.sub.1}
\includegraphics[width=0.40\textwidth]{{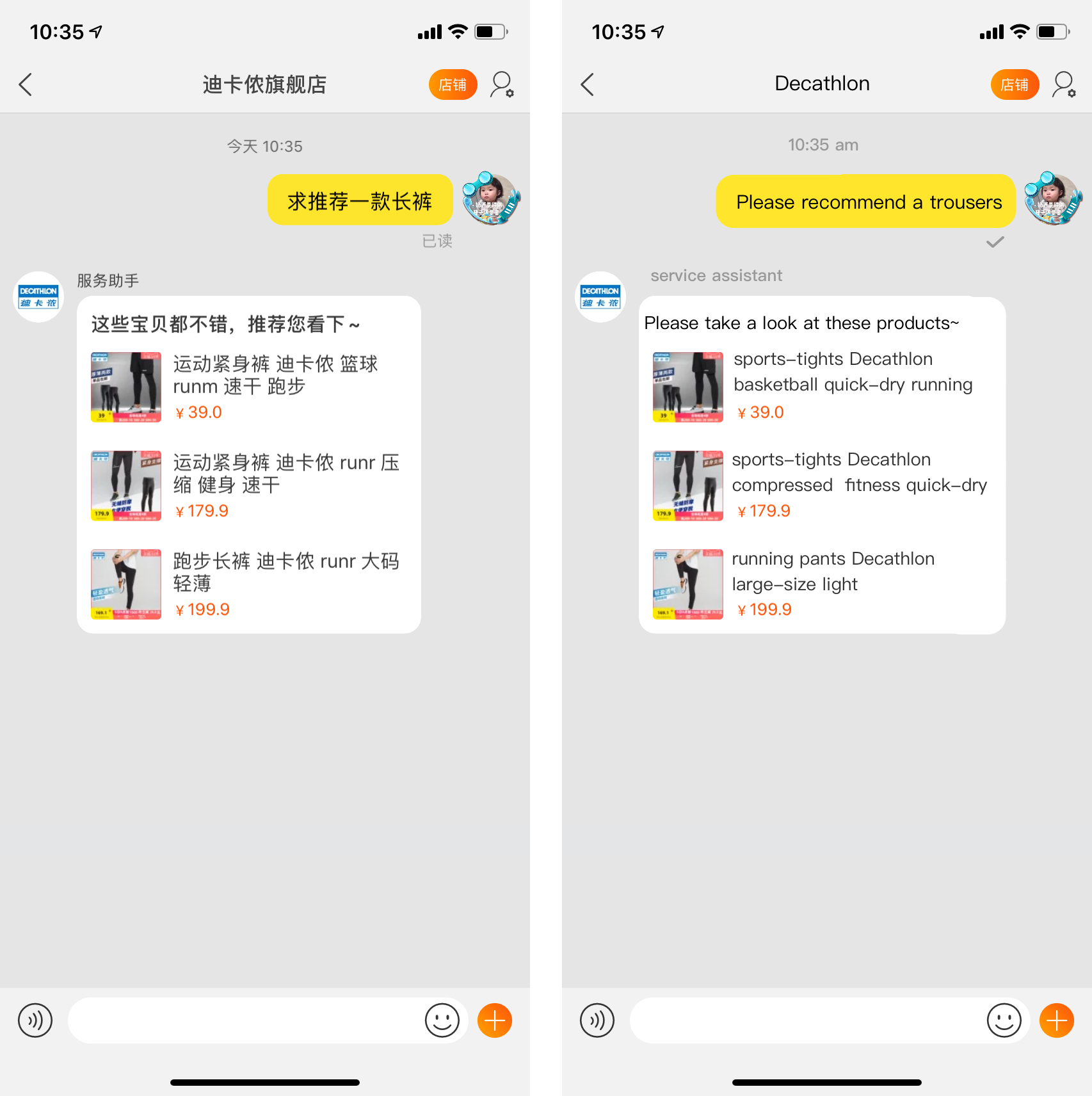}}}
\subfigure[example B]{
\label{Fig.sub.2}
\includegraphics[width=0.40\textwidth]{{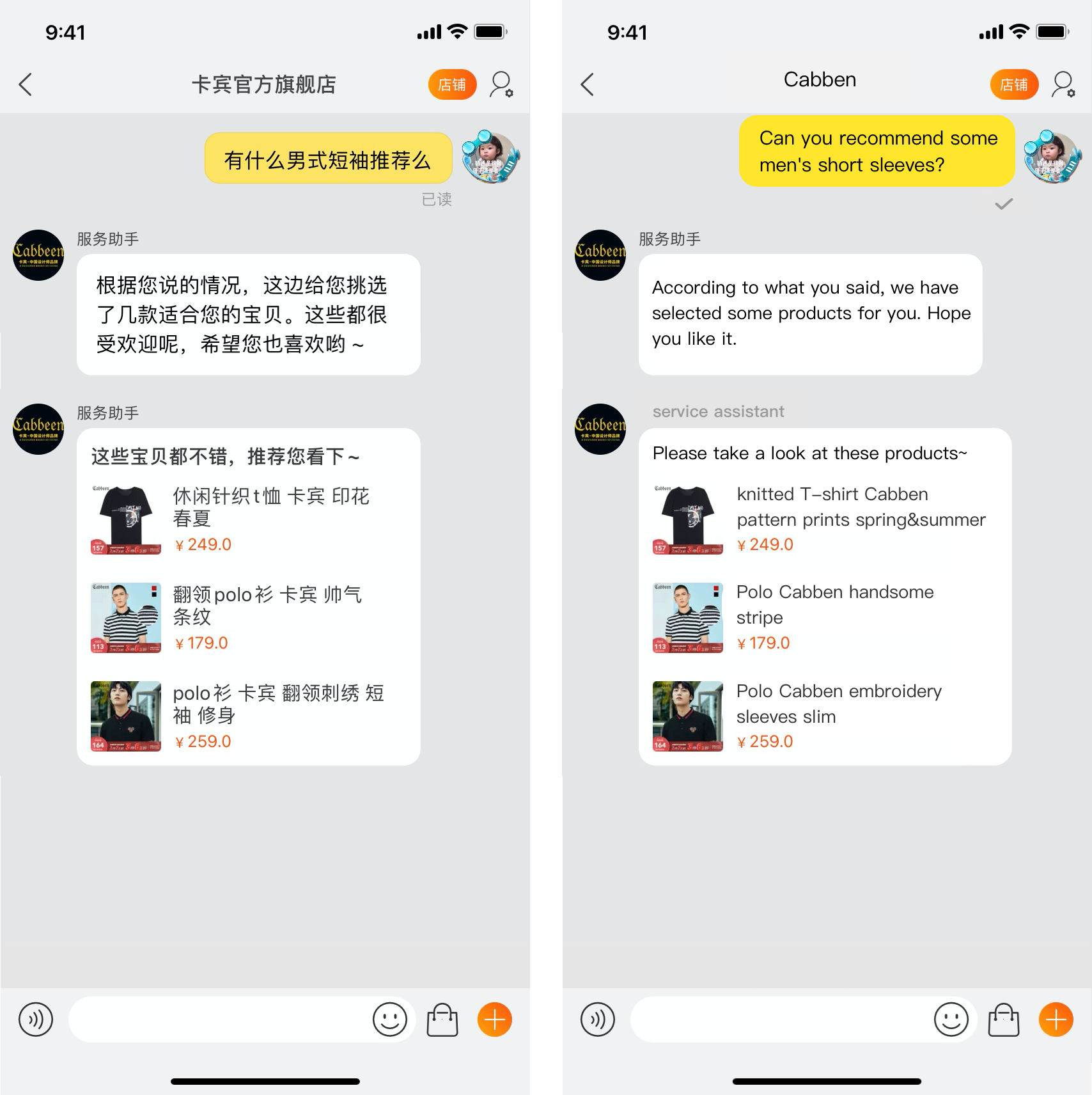}}}
\caption  {Two Real-word examples of conversation from Taobao with English translation on the right.}
\label{Fig.main}
\end{figure*}

\section{Experiment and Discussion}
\subsection{Datasets} K-DCN is pretrained and fine-tuned on a private dataset from scratch. The dataset for training CTR prediction model is from the conversation of Alime salebot (As illustrated in Figure 1).  As shown in table \ref{stat},We have sampled 900K records from 10 categories. 

\subsection{Implementation Details}
\begin{itemize}
    \item CKG. The pre-trained KG model is trained on  billion-scale converation knowledge graph (CKG). CKG has a total number of 500M entities and 6B triplets. We processd the data into triplets using Alibaba max-reduce framework (called Maxcompute). We removed the attributes with occurrences that are less than 5000 in CKG. Such attributues  are very sparse that are likely to deteriorate the model performance. For trainging, we employed the tool Graph-learn from \cite{zhu2019aligraph} to perform sampling of 10 neighbors for structure pretrain, and the batch size is 512. Adam optimizer is employed with initial learning rate = 0.0001; each training batch size = 1000, and node embedding size = 64. The model is trained with 50 parameter servers and 200 GPUs for 5 epochs. The whole training consumed 2 days.

    \item K-DCN. The K-DCN is implementated on Tensorflow.
The hyperparameters are tuned with grid search.
The optimal hyperparameter
settings were 2 deep layers of size 512 and 4 cross layers for
the K-DCN model.
More specifically, we use 4 kinds of behaviors in user state representation and the size of 
the convolutional operator is $N\times itemEmbeddingSize$, where $N$ is 2 and 4.
And the attention head number for text embedding is 4.
For different size of convolutional operator, the outputs are concated as final output. We use mini-batch stochastic optimization with
Adam \cite{kingma2014adam} optimizer and the batch size is set at 512.   
\end{itemize}




\subsection{Results and Discussion} 
We compare our methods with 5 baselines: Deep \& Cross Network (DCN) \cite{wang2017deep} , Wide\&Deep Network (WDN)\cite{cheng2016wide}, Deep Neural Network (DNN) \cite{liu2017survey}, Gradient Boosted Decision Trees (GBDT)\cite{ke2017lightgbm}, and Logistic Regression (LR) \cite{kleinbaum2002logistic} in 10 datasets, and the results are shown in \ref{result}. 
The proposed K-DCN outperforms all the baselines in all datasets. 

\begin{figure}[htbp]
  \center{\includegraphics[width=8.5cm,height=6cm]
  {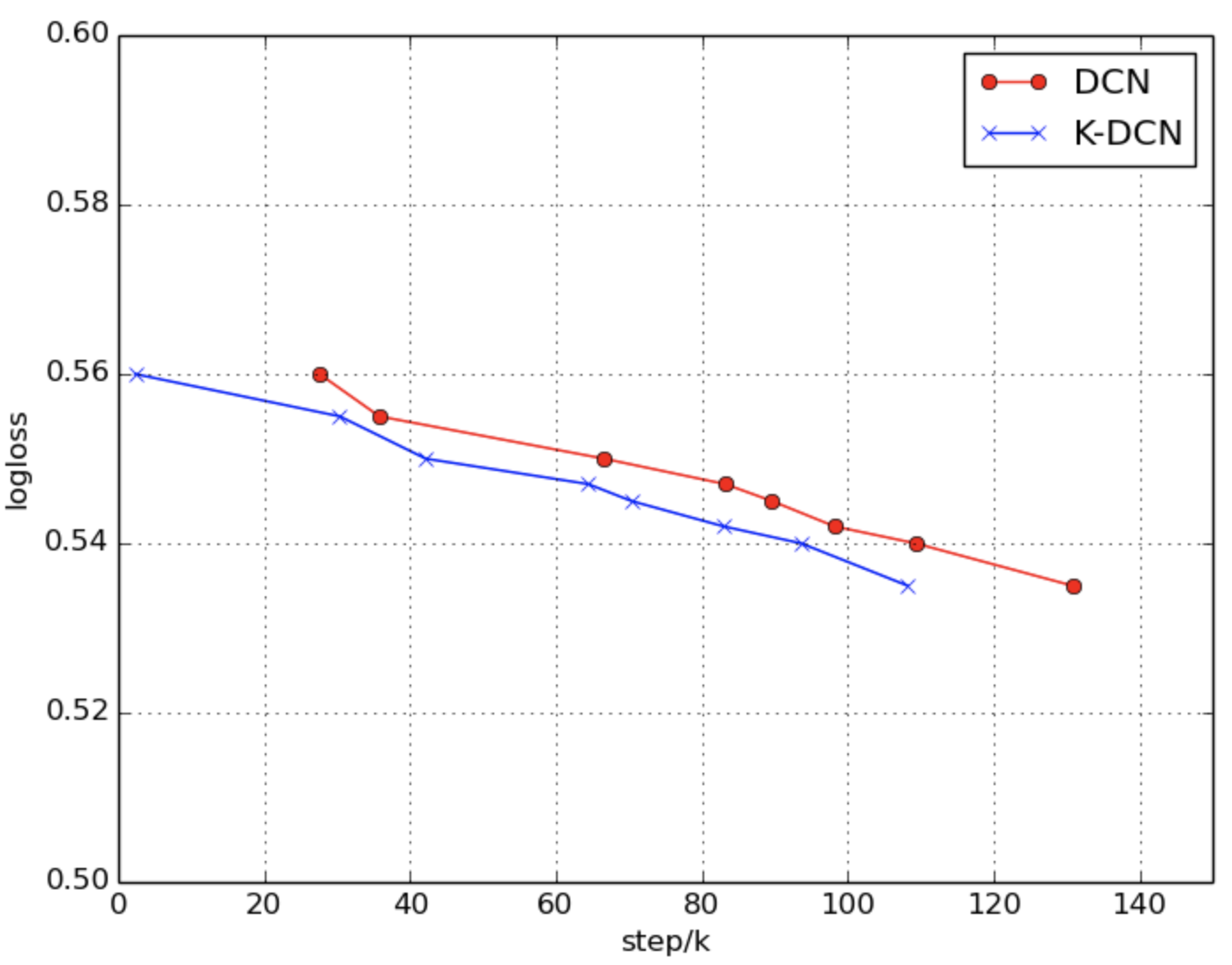}}
  \caption{Convergence speed comparison between K-DCN and DCN.}
  \label{conver}
\end{figure}

Specifically, K-DCN achieves a higher score of 2.6\% for the digital accessories category. The main reason is that the number of training data is very limited for digital accessories category, which verifies our model which was pretrained from conversation knowledge can alleviate the issue of data scarcity. For all datasets, K-DCN has an average improvement of 1.2\%, which verifies that the proposed KG embeddings from conversation knowledge graph can provide meaningful and useful structure and semantic information to CTR prediction model.  What's more, as shown in Figure \ref{conver}, it would also accelerate the convergence when the pretrained KG embedding were used to initialize the CTR prediction model. We have also tested our K-DCN with DCN in real application in Alime (as shown in Figure. \ref{Fig.main}. From table \ref{abtest}, K-DCN has relative improvement of 4.2\% over DCN, which proved the effectiveness of K-DCN.


\begin{table}[htbp]
\centering
 \caption{\label{tab:test} The number of training and testing data for deep CTR prediction model.}
  \begin{tabular}{|c|c|c|}
  \hline
  \textbf{Category} & \textbf{\# Train} & \textbf{\# Test}  \\ \hline
  Beauty Skin Care/Body Care/Essential Oil(BS) & 157027 &	16028\\ \hline
Women's wear(MW) &139167	&14205\\ \hline
Sports shoes(SS)    &104182	&10634\\ \hline
Women's and Men's underwear/Home wear(WN)
&98705	&10075\\ \hline
Sports clothes/Casual wear(SC) &93180	&9511\\ \hline
Bed linings (BL)&87537	&8935\\ \hline
Cleanser/Sanitary Napkin/Paper/Aromatherapy(CS)&58410	&5962\\ \hline
Men's clothes(MC)&57263	&5845\\ \hline
Digital accessories(DA)&49455	& 5048\\ \hline
Kitchen appliances(KA)&55069	&5621\\ \hline
  \end{tabular}
 \label{stat}
  \end{table}

\begin{table}[htbp]
\centering
 \caption{\label{tab:test} K-DCN and 5 baselines' performance in 10 datasets.}
  \begin{tabular}{c|cc|cccc}
  \hline
          \multirow{2}{*}{\textbf{Category}}   & \multicolumn{6}{c}{\textbf{AUC}}                                                                              \\ \cline{2-7} 
                              & \textbf{K-DCN}                     & \textbf{DCN}                          & \textbf{WDL}   & \textbf{DNN}   & \textbf{GBDT}  & \textbf{LR}    \\ \hline
  BC & {\color[HTML]{333333} \textbf{0.617}} & {\color[HTML]{333333} 0.608} & 0.6 & 0.598 & 0.584 & 0.58 \\ \hline
WM & {\color[HTML]{333333} \textbf{0.652}} & {\color[HTML]{333333} 0.636} & 0.63 & 0.633 & 0.611 & 0.592 \\ \hline
SS & {\color[HTML]{333333} \textbf{0.658}} & {\color[HTML]{333333} 0.652}  & 0.65  & 0.649  & 0.619  & 0.603  \\ \hline
WN & {\color[HTML]{333333} \textbf{0.663}} & {\color[HTML]{333333} 0.65} & 0.647 & 0.646 & 0.616 & 0.594 \\ \hline
SC & {\color[HTML]{333333} \textbf{0.667}} & {\color[HTML]{333333} 0.66} & 0.653 & 0.65 & 0.618 & 0.604 \\ \hline
BL & {\color[HTML]{333333} \textbf{0.631}} & {\color[HTML]{333333} 0.62} & 0.61 & 0.604 & 0.598 & 0.574 \\ \hline
CS & {\color[HTML]{333333} \textbf{0.625}} & {\color[HTML]{333333} 0.624} & 0.622 & 0.621 & 0.61 & 0.602 \\ \hline
MC & {\color[HTML]{333333} \textbf{0.647}} & {\color[HTML]{333333} 0.63} & 0.628 & 0.626 & 0.603 & 0.574 \\ \hline
DA & {\color[HTML]{333333} \textbf{0.706}} & {\color[HTML]{333333} 0.68} & 0.677 & 0.671 & 0.675 & 0.646 \\ \hline
KA & {\color[HTML]{333333} \textbf{0.654}} & {\color[HTML]{333333} 0.64} & 0.642 & 0.64 & 0.638 & 0.622 \\ \hline
  \textbf{Average} & \textbf{0.652}&0.64&0.636&0.633&0.617&0.59 \\ \hline
  \end{tabular}
 \label{result}
  \end{table}

\begin{table}[htbp]
\centering
 \caption{\label{tab:test} The online AB Test in Alime.}
  \begin{tabular}{|c|c|}
  \hline
  \textbf{Platform} & \textbf{\% Improvement of K-DCN over DCN}  \\ \hline
  Alime &4.2\% \\ \hline
  \end{tabular}
 \label{abtest}
  \end{table}

\section{Conclusion}
In this paper, we put forward a two-step CTR prediction model K-DCN.
We first propose a pretrained model KGM based on a billions scale conversation knowledge graph. 
Then we propose a knowledge-enhanced deep cross network (K-DCN) model based on KGM as well as some dense and sparse features. Experimental results show that our model obtains substantial performance on 10 datasets compared to baselines.
K-DCN have been applied to a real conservational chatbot in one of the largest E-commerce company.
In future work, we plan to design a learning framework which can uniformly exploit evidence from heterogeneous knowledge sources, such as entity descriptions and knowledge schema.


\bibliographystyle{IEEEtran}
\bibliography{conference_101719}

\end{document}